\documentclass[aps,prl,twocolumn,superscriptaddress,showpacs]{revtex4}

\usepackage{graphicx}
\usepackage{amsmath}

\begin{document}

\title{Effective Random Matrix Theory description of chaotic 
Andreev billiards}

\author{A. Korm\'anyos}
\affiliation{Department of Physics of Complex Systems, E{\"o}tv{\"o}s
University, H-1117 Budapest, P\'azm\'any P{\'e}ter s{\'e}t\'any 1/A, Hungary}
\author{Z. Kaufmann}
\affiliation{Department of Physics of Complex Systems, E{\"o}tv{\"o}s
University, H-1117 Budapest, P\'azm\'any P{\'e}ter s{\'e}t\'any 1/A, Hungary}
\author{C.~J. Lambert}
\email{c.lambert@lancaster.ac.uk}
\affiliation{Department of Physics, Lancaster University, Lancaster,
LA1 4YB, UK}
\author{J. Cserti}
\email{cserti@galahad.elte.hu}
\affiliation{Department of Physics of Complex Systems, E{\"o}tv{\"o}s
University, H-1117 Budapest, P\'azm\'any P{\'e}ter s{\'e}t\'any 1/A, Hungary}


\begin{abstract}

An effective random matrix theory description is developed for the
universal gap fluctuations and the ensemble averaged density of
states of chaotic Andreev billiards for finite Ehrenfest time.  It
yields a very  good agreement with the numerical calculation for
Sinai-Andreev billiards. A systematic linear decrease of the mean
field gap with increasing Ehrenfest time $\tau_E$ is observed but
its derivative with respect to $\tau_E$ is in between two
competing theoretical predictions and close to that of the recent
numerical calculations for Andreev map. The exponential tail of
the density of states is interpreted semi-classically.

\end{abstract}


\maketitle

Recently, mesoscopic ballistic two dimensional normal (N) dots in
contact with a superconductor(S) have been extensively studied.
Such hybrid systems are commonly called Andreev
billiards\cite{Kosztin,Melsen,Altland,Heny,Ihra}. In the most
recent works, interest has shifted to mesoscopic fluctuations of
the excitation spectrum of these systems
\cite{Univ-Gap-ref,kicked-rotator,Ostrovsky,Goorden}. Since the
sub-gap spectrum determines the tunneling conductance of an N-S
contact this is an essential question both experimentally and
theoretically.

Based on the semi-classical treatments and random matrix theory
(RMT), it was shown by Melsen et al.~\cite{Melsen} that integrable
Andreev billiards are gapless, whereas  systems with classically
chaotic dots possess an energy gap on the scale of the Thouless
energy $E_T=\hbar/(2\tau_D)$, where $\tau_D = \pi A/(W v_{\rm F})$
is the mean dwell time in the normal dot (here $A$ is the area of
the normal dot, $W$ is the width of the superconducting region and
$v_{\rm F}$ is the Fermi velocity). For such systems, it is
assumed that $\delta_N << E_T << \Delta$, where $\delta_N=2\pi
\hbar^2/(mA)$ is the mean level spacing of the isolated normal dot
with effective mass $m$ of the electrons and $\Delta$ is the bulk
order parameter of the superconductor\cite{Heny}. In further
studies\cite{Univ-Gap-ref,Ostrovsky} it was concluded that in
chaotic cases, the lowest energy level $E_1$ of the system varies
from sample to sample with a universal probability distribution
$P(x)$ given in Ref.~\onlinecite{Univ-Gap-ref} if the energy
levels $E_1$ are rescaled as $x=(E_1 - E_g)/\Delta_g$, where the
mean-field value of the gap $E_g$ and the width of the
distribution $\Delta_g$ are given by
\begin{subequations}
\begin{eqnarray}
E_g &=& E_g^{\rm {\scriptscriptstyle RMT}} = 2\gamma^{5/2} E_T,  \\
\label{Eg-RMT:def}
\Delta_g &=& \Delta_g^{\rm {\scriptscriptstyle RMT}} =
c^\prime M^{1/3}\delta_N.
\label{Dg-RMT:def}
\end{eqnarray}
\label{Eg-Dg-RMT:def}
\end{subequations}
Here $\gamma = \frac{1}{2}(\sqrt{5}-1)$ is the golden ratio ,
$c^\prime~= {\left[\left(15-6\sqrt{5} \right)/20\right]}^{1/3}/2\pi$,
$M=\rm{Int}\left[k_{\rm F}W/\pi\right]$
is the number of open channels in the S region and
$k_{\rm F}$ is the Fermi wave number
($\rm{Int}[.]$ stands for the integer part).

Equations (\ref{Eg-Dg-RMT:def}) are strictly valid only in the RMT
limit, i.e., when the Ehrenfest time $\tau_{\rm E} = (1/\lambda
)\, \ln (L/\lambda_{\rm F})$ tends to zero ($\tau_{\rm E}$  is the
time needed for a wave packet of minimal size $\lambda_{\rm
F}=2\pi/k_{\rm F}$ to spread to the characteristic size $L$ of the
classically chaotic normal dot with Lyapunov exponent $\lambda$).
For finite but small enough Ehrenfest time, Silvestrov et al.\
\cite{Silvestrov}, and Vavilov and Larkin\cite{Vavilov-Larkin}
predicted that to lowest order in $\tau_{\rm E}/\tau_D$ the
mean-field gap $E_g$ decreases linearly by increasing the ratio
$\tau_{\rm E}/\tau_D$. The first numerical evidence for the
distribution $P(x)$ in the RMT limit and the dependence of $E_g$
on the ratio $\tau_{\rm E}/\tau_D$ was presented by Jacquod et
al.~\cite{kicked-rotator} modeling the hybrid system with the {\em
one dimensional} Andreev map.

From an experimental point of view, more realistic candidates for
studying quantum chaos in hybrid systems would be two dimensional
Andreev billiards with classically chaotic normal region.
\begin{figure}[hbt]
\includegraphics[scale=0.35]{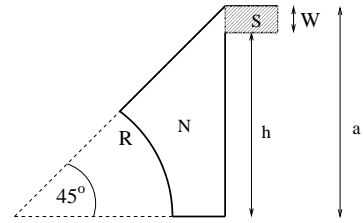}
\caption{A normal dot (N) of Sinai billiard in contact with a
superconductor (S).
\label{geometria-fig}}
\end{figure}
However, to date no numerical study of these systems confirms the
predictions of the RMT, except for the mean field density of
states\cite{Melsen}. The purpose of this paper is twofold. On the
one hand, we present (for the first time) a numerical study of the
gap fluctuation and the density of states (DOS) in {\em two
dimensional} chaotic Andreev billiards. On the other hand, to show
that for finite Ehrenfest time the gap distribution $P(x)$ is
still the {\em same} universal function given in the RMT limit but
the scaling parameters $E_g$ and $\Delta_g$ extracted from the
ensemble (see below) are different from those given in
Eq.~(\ref{Eg-Dg-RMT:def}). In this way, this is an  {\em
effective} RMT description of the fluctuation of the gap using
`renormalized' scaling parameters. As a further support of our
model it is also shown that this effective RMT description
correctly predicts  the edge of the ensemble averaged DOS. For
clarity, we would like to mention that the recently developed
theory\cite{Silvestrov} for the mean field gap based on the
concept of a reduced phase space can also be considered as an
effective RMT description. However, our effective RMT description
deals with the fluctuation of the gap in Andreev billiards.

In our numerical work we calculated the {\em exact} energy levels of
the so-called Sinai-Andreev (SA) billiard in which the normal dot is
a Sinai billiard (see Fig.~\ref{geometria-fig}).
The energy levels of the Andreev billiards  are the positive
eigenvalues $E$ (measured from the Fermi energy) of the
Bogoliubov-de Gennes equation\cite{deGennes-book}.
To obtain the exact energy levels of an Andreev billiard
we used the recently derived general and
quantum mechanically exact secular
equation expressed in terms of the scattering matrix $S_0(E)$
of the normal region\cite{Gap-cikk}.
The scattering matrix $S_0(E)$ was calculated by expanding the wave
function in the N region in terms of Bessel functions.

To ensure that the long classical trajectories
(compared to the characteristic length of the system)
starting and ending at the N-S
interface are truly chaotic the following geometrical constrains
should be applied: $h=a-W$ (the superconductor is placed
at the top of the vertical border of the Sinai billiard),
$R+W \ge a$ and $R \ge a/\sqrt{2}$.
Otherwise, there may exist arbitrary long trajectories
without bouncing on the circular part of the Sinai billiard.
For such intermittent trajectories the return probability
decays as $P_r(s)\sim 1/s^3$ ($s$ is the length of the trajectory)
and this results in a gapless energy spectrum of the
Andreev billiard\cite{Melsen,Heny}.

For realistic systems the Ehrenfest time is always finite and an
important question arises, namely how the predictions of the RMT
for the gap distribution and  the scaling parameters $E_g$ and
$\Delta_g$ are affected. Provided that for small enough $\tau_{\rm
E}/\tau_D$ the distribution of the gap in the rescaled variable
$x$ remains the {\em same} as that in the RMT limit, the following
equations hold
\begin{subequations}
\begin{eqnarray}
E_g &=& \langle E_1 \rangle - \langle x \rangle \Delta_g,
\label{Eg-atlag:def}    \\
\Delta_g &=& \delta E_1/\delta x,
\label{Dg-atlag:def}
\end{eqnarray}
\label{Eg-Dg-atlag:def}
\end{subequations}
where $\langle E_1 \rangle$ is the mean value,
$\delta E_1 = \sqrt{\langle E_1^2 \rangle - \langle E_1 \rangle^2}$
is the standard deviation of $E_1$, while
$\langle x \rangle \approx 1.21$ and
$\delta x = \sqrt{\langle x^2\rangle - \langle x \rangle^2} \approx 1.27$
are calculated from the gap distribution $P(x)$\cite{Univ-Gap-ref}.
Although the distribution of $E_1$ is unknown, its mean and standard
deviation can be numerically estimated from the data of an ensemble of
the SA billiard, hence $E_g$ and $\Delta_g$ follow.

\begin{figure}[hbt]
\includegraphics[scale=0.47]{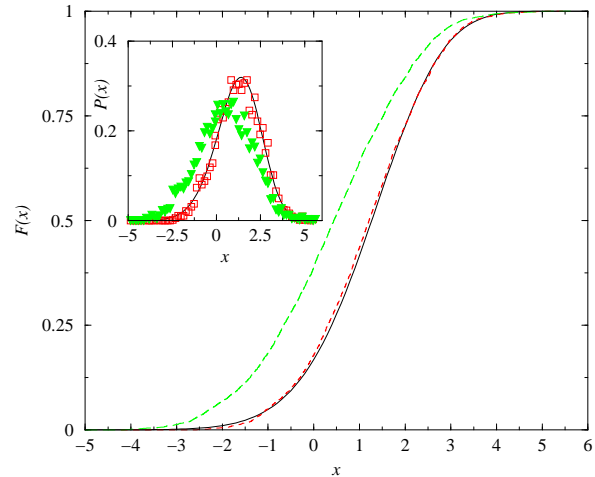}
\vspace*{-5mm}
\caption{Comparison of our numerical result
using the parameters $E_g$ and $\Delta_g$ given by
Eq.~(\ref{Eg-Dg-atlag:def}) (short dashes) and
Eq.~(\ref{Eg-Dg-RMT:def}) (long dashes)
with the theoretical predictions of
the integrated distribution $F(x)$ (solid line).
In the inset the same as in
the main frame is plotted for the distribution $P(x)$ with square and
triangular symbols, and solid line, respectively.
\label{fig-Fx}}
\end{figure}

Figure \ref{fig-Fx} shows our numerical results for the integrated
distribution $F(x)= \int_0^x P(x^\prime)\, d x^\prime$ together
with the theoretical prediction (the distribution $P(x)$ is shown
in the inset). In our numerics we used 5000 slightly different
realizations of the SA billiard by varying the geometrical
parameters $R$, $W$ and the Fermi wave numbers $k_F$ (for the
parameters of the SA billiard see\cite{details_ensemble-1}). From
Eq.~(\ref{Eg-Dg-atlag:def}) we found that $E_g\approx0.5\,E_T$ and
$\Delta_g \approx0.118\,E_T $ and they are different from those
given in the RMT limit, $E_g^{\rm {\scriptscriptstyle RMT}} =
0.6\,E_T$ and $\Delta_g^{\rm {\scriptscriptstyle RMT}} =
0.097\,E_T$. It is clear from the figure that the numerical result
for $F(x)$ (using Eq.~(\ref{Eg-Dg-RMT:def})) is also different
from that of the theoretical prediction in the RMT limit. However,
the agreement between our numerical results obtained using
Eq.~(\ref{Eg-Dg-atlag:def}) and the universal distribution
function $F(x)$ is excellent, without any adjustable parameters.
Similar numerical results were found for other ensembles of the SA billiard.
These results imply that for systems with non-zero Ehrenfest time
the gap distribution is still given by the universal function $P(x)$
provided the `renormalized' parameters $E_g$ and $\Delta_g$ obtained from
Eq.~(\ref{Eg-Dg-atlag:def}) are used.
One can regard the distribution $P(x)$
(and the integrated distribution $F(x)$) with the renormalized parameters
as a result of an effective RMT description.

The necessity of the renormalization of $E_g$ and $\Delta_g$ can
be interpreted as their dependence on the Ehrenfest time. Since
the characteristic length of the system is uncertain, two
definitions of the Ehrenfest time, proposed in
Ref.~\onlinecite{Vavilov-Larkin} and used in numerical simulations
of Ref.~\onlinecite{kicked-rotator}, were here adapted for
numerical calculations:
\begin{subequations}
\begin{eqnarray}
\tau_{\rm E}^{\left(1\right)}
&=& \frac{1}{2\lambda}\, \ln \frac{W^2}{\lambda_{\rm F}L_c},
\label{Eh-1}  \\[1ex]
\tau_{\rm E}^{\left(2\right)}
&=& \frac{1}{2\lambda}\, \ln \frac{L_{\rm av}}{\lambda_{\rm F}},
\label{Eh-2:def}
\end{eqnarray}
\label{Eh-12:def}
\end{subequations}
where $L_c$ is the average length of the part of the trajectory
lying between two consecutive bounces at the curved boundary
segment of the Sinai billiard and $L_{\rm av} = \pi A/K $ with
perimeter $K$ of the billiard is the mean chord length in the
normal region. They are parametrically different, but their
numerical values are of the same magnitude. For the numerical
results shown in Fig.~\ref{fig-Fx} the ratio of the Ehrenfest time
and the dwell time is about $\tau_{\rm E}^{\left(1\right)}/\tau_D
\approx 0.1$ and $\tau_{\rm E}^{\left(2\right)}/\tau_D \approx
0.26$ (for calculation of the Lyapunov exponent
see\cite{details_ensemble-1}).

It was predicted theoretically\cite{Silvestrov,Vavilov-Larkin}
and demonstrated numerically using the Andreev map\cite{kicked-rotator}
that to lowest order in $\tau_{\rm E}/\tau_D$
the mean field gap $E_g$  decreases linearly by
increasing the ratio of the Ehrenfest time and the dwell time for ratio
much less than one.
\begin{figure}[hbt]
\includegraphics[scale=0.4]{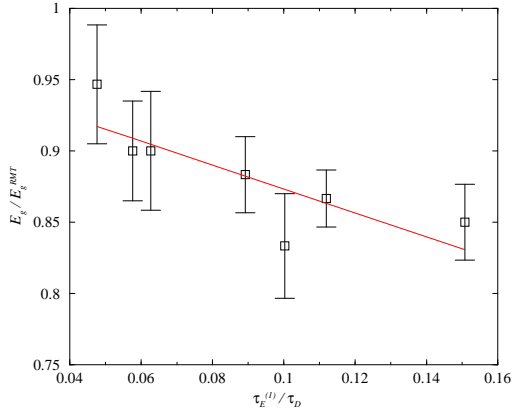}
\vspace*{-3mm}
\caption{The mean field gap $E_g/E_g^{\rm {\scriptscriptstyle RMT}}$
as a function of ratio $\tau_{\rm E}^{\left(1\right)}/\tau_D$.
The straight line is the numerical fit using Eq.~(\ref{slope:eq}).
\label{fig-Eg}}
\end{figure}
From our numerical study a systematic decrease of $E_g$
as a function of ratio $\tau_{\rm E}^{\left(1\right)}/\tau_D$ is found
as shown in Fig.~\ref{fig-Eg} (for details of the system parameters
see\cite{details_ensemble-2}).
Similar result has been obtained using
definition (\ref{Eh-2:def}) of the Ehrenfest time.
The error bars represent the standard deviations which
are calculated by taking into account the correlation
between the first energy levels $E_1$ for systems with
slightly different parameters (see\cite{error_bar} for details).

Assuming that
\begin{equation}
\frac{E_g}{E_g^{\rm {\scriptscriptstyle RMT}}} = \beta
- \alpha \, \frac{\tau_{\rm E}^{\left(1\right)}}{\tau_D},
\label{slope:eq}
\end{equation}
 we found
that $\alpha = 0.7 \pm 0.2$ and $\beta = 0.95 \pm 0.02$.
Similar results were obtained by using (\ref{Eh-2:def}):
$\alpha = 0.9 \pm 0.3$ and $\beta = 1.10 \pm 0.07$.
According to the two competing theories the universal values of
$\alpha =0.23$ and $\beta= 1$ (Ref.~\onlinecite{Vavilov-Larkin}) and
$\alpha =2$ and $\beta= 1$ (Ref.~\onlinecite{Silvestrov}) were predicted,
while from the numerics for Andreev map\cite{kicked-rotator} only
$\alpha = 0.59\pm 0.08$ is  universal value.
From our numerics the value $\alpha$ is in between the two
theoretical predictions  and within the numerical errors it agrees
with the result for the Andreev map.
We found that the values of $\Delta_g$ obtained from
(\ref{Dg-atlag:def}) are slightly greater than those predicted from
(\ref{Dg-RMT:def}). However, no obvious functional form
can be deduced from our data
for the dependence of $\Delta_g$ on $\tau_{\rm E}/\tau_D$.

\begin{figure}[hbt]
\includegraphics[scale=0.47]{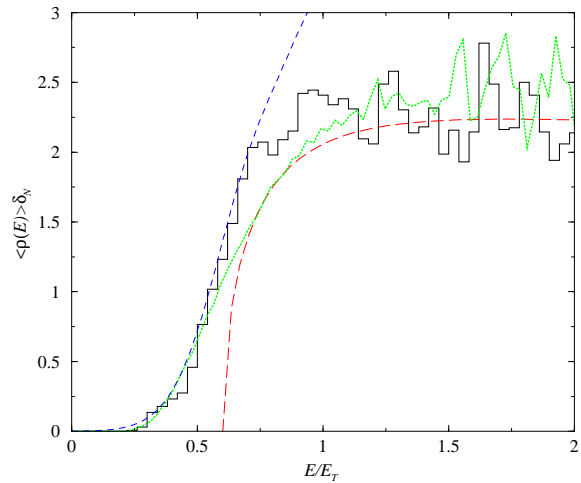}
\vspace*{-4mm}
\caption{Ensemble averaged DOS $\langle \varrho(E) \rangle$
(multiplied by $\delta_N$) as a function of $E/E_T$ for SA
billiard (solid line), the effective RMT prediction
$\langle \varrho_{\rm eff}(E) \rangle$
(expressing $x$ in terms of $E$ by using Eq.~(\ref{Eg-Dg-atlag:def}))
(short dashes),
the Bohr-Sommerfeld calculation (dotted line),
and the mean field DOS (long dashes).
\label{fig-DOS-E_Th}}
\end{figure}

For further support of  the effective RMT description of
the universal gap fluctuation we compare the numerically obtained
ensemble averaged DOS
$\langle \varrho(E) \rangle$ with the effective RMT prediction.
The ensemble averaged DOS
$\langle \varrho_{\rm eff}(E) \rangle$
in the effective RMT description can be calculated from
$\langle \varrho_{\rm {\scriptscriptstyle RMT}}(x) \rangle$
given in the RMT limit using the renormalized parameters $E_g$ and
$\Delta_g$ when `scaling back' the variable $x=(E-E_g)/\Delta_g$ into
the energy variable $E$.
In the RMT limit
$\langle \varrho_{\rm {\scriptscriptstyle RMT}}(x) \rangle =
-x{\rm Ai}^2(x)+{\left[{\rm Ai}^\prime(x)\right]}^2 +
\frac{1}{2}\,{\rm Ai}(x)\,
\left[1-\int_x^\infty \,{\rm Ai}(y)\, dy \right]$ is again
a universal function of $x$ (see note 20 in \cite{Univ-Gap-ref}).
It can be seen from Fig.~\ref{fig-DOS-E_Th} that
the agreement between our numerically obtained
ensemble averaged DOS $\langle \varrho(E) \rangle$
and $\langle \varrho_{\rm eff}(E) \rangle$ is excellent  at the edge of
the spectrum (where the theory is valid)
for ratio $\tau_{\rm E}^{\left(1\right)}/\tau_D \approx 0.063$
(the same holds for other ratios not shown).
For larger energies  the DOS is around $2/\delta_N$
as it is expected.
For convenience the mean field DOS\cite{Melsen}
in the RMT limit is also plotted in the figure.
In case of two dimensional Andreev billiards the ensemble averaged
DOS $\langle \varrho(E) \rangle$ shown
in Fig.~\ref{fig-DOS-E_Th} is the first
numerical evidence
for the prediction $\langle \varrho_{\rm eff}(E) \rangle$
obtained from the theoretical result
$\langle \varrho_{\rm {\scriptscriptstyle RMT}}(x) \rangle$
(see Fig. 2 of Ref.~\cite{Univ-Gap-ref}).

From Fig.~\ref{fig-DOS-E_Th} one can also see that the agreement
between the Bohr-Sommerfeld approximation (BS)\cite{Melsen} and
our numerics  is quite good at the bottom of the spectrum. This
suggests a semi-classical explanation. In a work by Schomerus and
Beenakker\cite{Heny} close correspondence has been found between
the morphology of the phase space and the density of low energy
excitations. Ref.~\cite{Heny} finds that the semi-classical
prediction of the DOS for systems with fully chaotic phase space
has no definite gap but it becomes exponentially small below the
energy $\approx 0.5 E_{\rm T}$. Furthermore in case of systems
with mixed phase space and strong coupling of the regular islands
to the superconductor, the above defined `gap' is substantially
reduced, namely by a factor $\tau_{D}/t^*$, where $t^*$ is the
mean dwell time of trajectories in the chaotic part of the phase
space.

The concept of the gap reduction can also be applied to our SA
billiard, since this is a system with mixed phase space (for
geometries studied in this work approximately 10\% of the phase
space was regular). To calculate the mean dwell time $t^*$ in the
chaotic part of the phase space, only those trajectories are taken
into account, for which the size of the bunch of trajectories
started in its $\lambda_{\rm F}$ neighborhood at the N-S interface
is increased to the characteristic length scale of the billiard
before returning to the superconducting lead. For the parameters
 of the SA billiard used in our calculations
this means that one has to exclude trajectories which
either do not bounce on the circular part  or bounce on it only once.
From our numerics we found  (for the ensemble corresponding
to Fig.~\ref{fig-DOS-E_Th}  that
the effective dwell time of these trajectories is
$t^*  \approx 1.66 \tau_{\rm D}$ hence the semi-classically obtained
gap is reduced to the value  $\approx \! 0.3E_{\rm T}$.
This result is just about the energy value where the numerically found
DOS becomes exponentially small and thus it
confirms the semi-classical picture developed in Ref.~\cite{Heny}.
Note also that below the value  $\approx \! 0.3E_{\rm T}$
(i.e., for $x\lesssim -2$) the RMT DOS
$\langle \varrho_{\rm {\scriptscriptstyle RMT}}(x) \rangle$ is also
exponentially small.

While the BS approximation is quite successful in predicting the
density of low energy excitations, it is clear from
Fig.~\ref{fig-DOS-E_Th}  that the edge of the spectrum can be
better predicted using our effective RMT description. This  may
also reveal the limits of the BS approximation. (We observed
similar deviations for other ensembles not shown).

In summary we have numerically shown that for finite but small enough
Ehrenfest time the distribution of the rescaled first energy level $E_1$
of an ensemble of chaotic Andreev billiards
can be treated by an effective RMT description.
In this model the scaling parameters $E_g$ and $\Delta_g$  extracted
from the data of the ensemble rescale the distribution of $E_1$
such that it agrees with $P(x)$ given in the RMT limit.
Our numerical results also show that to lowest order
in $\tau_{\rm E}/\tau_D$  the mean field gap $E_g$ decreases linearly with
the Ehrenfest time but the slope is between the two competing theories
and close to that of the recent numerical calculations for the Andreev map.
Calculation of the ensemble averaged DOS gives
further confirmation of our effective RMT description.
Our numerics suggest that the exponential tail of the DOS
can be well interpreted semi-classically.

We gratefully acknowledge very helpful discussions with
C.~W.~J.~Beenakker and M. G. Vavilov.
This work was supported in part by the European Community's Human Potential
Programme under Contract No. HPRN-CT-2000-00144, Nanoscale Dynamics,
the Hungarian-British Intergovernmental Agreement on Cooperation in
Education, Culture, and Science and Technology,
and the Hungarian  Science Foundation OTKA  T034832 and T029552.
One of us (Z. K.)  thanks the Hungarian Academy of Sciences
for its financial support as a J\'anos Bolyai Scholarship.


\begin{thebibliography}{10}
\bibitem{details_ensemble-1}

In our numerical calculations the bulk superconductor gap $\Delta$
was fixed: $\Delta = 0.02 E_{\rm F}$.
The width $W$ of the S region varies in $[0.19a,0.191a]$,
$R$ in the range $\left[0.868a,0.874a \right]$
(this variation corresponds to $0.4\,\lambda_{\rm F}$).
Values of $E_{\rm F}$ were chosen between
$E_{\rm F}^0 \pm 0.004E_{\rm F}^0$
with such an $E_{\rm F}^0$ that
the number of open channels in the S region is always $M=25$.
The Lyapunov exponent $\lambda$ was obtained by following
trajectories from $4\cdot 10^6$ uniformly distributed initial conditions.
For the above given range in $R$ the Lyapunov exponent slightly
changes in the range $\lambda ~= \left[1.65, 1.68\right]$.
\bibitem{details_ensemble-2}
In Fig.~\ref{fig-Eg} the number of open channels varies between $M=18$
and $30$, whereas the width $W$ between $0.16\, a$ and $0.22\, a$,
and the ranges in $R$ and $E_{\rm F}$
are the same as in \cite{details_ensemble-1}.
The number of realizations is between 2500 and 5000.
\bibitem{error_bar}

As can be seen from Eq.\ (2a) one needs to evaluate both the error
(i.e. standard deviation)
of the numerical approximation of $\langle E_1 \rangle$ and that
of $\Delta_g$.
It can be shown that both standard deviations are enhanced by
the correlations of $E_1$, thereby the error of $E_g$ is also effected.
The standard deviation of $\langle E_1 \rangle$ can be formulated as
$\sigma ( \langle E_1 \rangle) = \sqrt{\frac{N^*}{N}}\, \sigma (E_1)$,
where $N$ is the number of different realizations, $\sigma(E_1)$ is
the standard deviation of $E_1$, $N^*$ is a characteristic of the correlations.
In our studies $N^*$ was in the range $15 - 55$.
The error of $\Delta_g$ is estimated by
calculating $\Delta_g$ separately for parts of our ensemble and
analyzing its fluctuation.
\end{thebibliography}

\samepage{

}



\end{document}